\documentclass[preprint2]{aastex}

\shorttitle{H$_2$ in the Ionized Region of PNe}
\shortauthors{Aleman \& Gruenwald}

\begin{document}

\title{Molecular Hydrogen in the Ionized Region of Planetary Nebulae}
\author{Isabel Aleman\altaffilmark{1} and Ruth Gruenwald } 
\affil{Instituto de Astronomia, Geof\' \i sica e Ci\^{e}ncias Atmosf\'{e}ricas, 
Universidade de S\~{a}o Paulo, SP, Brazil}
\altaffiltext{1}{Send offprint requests to isabel@astro.iag.usp.br.}


\begin{abstract}

This paper presents an analysis of the concentration of the hydrogen molecule inside the ionized region of planetary nebulae. The equations corresponding to the ionization and chemical equilibria of H, H$^+$, H$^-$, H$_2$, H$_2^+$, and H$_3^+$ are coupled with the equations of ionization and thermal balance for a photoionized atomic gas. Forty different reactions related to the formation or the destruction of these species are included. The presence of dust is taken into account, since grains act as catalysts for the production of H$_2$, as well as shield the molecules against the stellar ionizing radiation. We analyze the effect of the stellar ionizing continuum, as well as of the gas and grain properties on the calculated H$_2$ mass. It is shown that a significant concentration of H$_2$ can survive inside the ionized region of planetary nebulae, mostly in the inner region of the recombination zone. The total H$_2$ to total hydrogen mass ratio inside the ionized region increases with the central star temperature, and, depending on the PN physical conditions, it can be of the order of $\sim$ 10$^{-6}$ or even higher. The increase of the recombination zone with the stellar temperature can account for such correlation. This can explain why the H$_2$ emission is more frequently observed in bipolar planetary nebulae (Gatley's rule), since this kind of object has typically hotter stars. Applying our results for the planetary nebula NGC 6720, we obtain an H$_2$ to hydrogen mass ratio similar to the value obtained from the observed H$_2$ line emission.

\end{abstract}


\section{INTRODUCTION}

Since the first detection of molecular hydrogen in NGC 7027 (Treffers et al. 1976), this molecule has been detected in many planetary nebulae (PNe), in both the infrared (Aspin et al. 1993; Kastner et al. 1996; Hora, Latter, \& Deutch 1999) and the ultraviolet (Bowers et al. 1995; Herald \& Bianchi 2002).

The analysis and interpretation of the infrared emission due to hydrogen molecules in PNe has been made in the context of: (a) dense neutral clumps inside the ionized region (Beckwith, Persson, \& Gatley 1978; Reay, Walton, \& Atherton 1988; Tielens 1993; Schild 1995), (b) photodissociation regions (PDRs; Tielens 1993; Natta \& Hollenbach 1998; Vicini et al. 1999), as well as (c) shocks between the expanding shell and the precursor red giant wind (Natta \& Hollenbach 1998). Visual inspection of H$_2$ and [NII] line images observed by several authors (Schild 1995; Guerrero et al. 2000; Arias et al. 2001; Bohigas 2001) can not exclude, however, that the H$_2$ emission is partly (or mostly) produced inside the ionized region. In fact, the presence of grains in these regions, the coexistence of electrons and hydrogen ions, and the corresponding thresholds for the destruction or ionization of the different species, suggest that H$_2$ molecules can survive inside ionized regions. Interpretation of the far ultraviolet absorption spectra recently obtained by the Far Ultraviolet Spectroscopic Explorer (FUSE) also suggests the existence of a hot component of excited H$_2$, which could be inside the ionized gas in PNe (Herald \& Bianchi 2002). 

Formation on grain surface is the main process for the production of H$_2$ molecules in the interstellar medium. This formation route was first proposed to explain the amount of H$_2$ found in interstellar clouds (Hollenbach \& Salpeter 1971). Grains are condensed in the atmospheres of asymptotic giant branch and red giant stars (see Whittet 1992 and Evans 1994) and, since these stars correspond to the earlier evolutionary stages of PNe, grains could be present in the ionized region if they are not destroyed by the stellar radiation. Continuum infrared radiation observations and modeling of PNe indicate that dust is present inside the ionized region of PNe (e.g., Natta \& Panagia 1981; Harrington, Monk, \& Clegg 1988; Lenzuni, Natta \& Panagia 1989; Stasinska \& Szczerba 1999; Van Hoof et al. 2000). In particular, the distribution of the infrared emission follows the morphology of infrared lines and continuum radiation emitted by the ionized gas (Barlow 1993; Meixner et al. 1996), also suggesting the presence of grains in the ionized region. A very strong evidence that dust is present in the ionized region comes from the models obtained by Kingdon, Ferland, \& Feibelman (1995) and Kingdon \& Ferland (1997), which show depletion of elements that can be present in grains. Some authors (for example, Stasinska 2002 and references therein) argue, however, that the grains are in clumps and condensations inside the ionized region. If grains are indeed mixed with the ionized gas, they can act as catalyst for the formation of H$_2$ molecules. 

The presence of electrons and/or H$^+$ inside the ionized gas can also produce hydrogen molecule nuclei, through radiative attachment and associative detachment reactions. On the other hand, the threshold for the photodissociation of hydrogen molecules is 14.7 eV, while the photoionization threshold is 15.4 eV. These values are higher than the threshold for the ionization of hydrogen atoms (13.6 eV). Hydrogen molecules could then be shielded from the photons that could dissociate it, surviving inside the ionized region in zones far from the ionizing source.

Detailed analyses of the presence of molecules have been done for interstellar clouds (for example, de Jong 1972; Black \& Dalgarno 1977; Viala, Roueff, \& Abgrall 1988), as well as for photodissociation regions (for example, Tielens \& Hollenbach 1985; Draine \& Bertoldi 1996). Models for photodissociation regions have been applied to planetary nebulae, assuming that the incident radiation is filtered after crossing the ionized region (e.g., Natta \& Hollenbach 1998). These models assume that the flux of the incidente radiation is zero between 13.6 and 54.4 eV. Due to the behavior of the ionizing cross-sections with the photon energy, the correct ionizing spectrum that reaches each point of the ionized region is, however, much more complex, and a proper account of the radiation transfer along the nebula must be done when analyzing the molecular chemistry.

Since the first detection of molecular hydrogen in PNe, only a few attempts to calculate the H$_2$ content in ionized regions have been made (Black 1978; Gussie \& Pritchet 1988; Cecchi-Pestellini \& Dalgarno 1993). In these studies, however, one or more of the following effects is not taken into account: the effect of dust on the radiation transfer, the formation of H$_2$ molecules on grain surface, and the photodissociation in two steps.

An interesting correlation has been obtained between the presence of H$_2$ line emission and the morphological type of the planetary nebula. Significant H$_2$ emission is more frequently observed in PNe having bipolar morphology (Gatley's rule; Zuckerman \& Gatley 1988; Kastner at al. 1996). This correlation between H$_2$ emission and the morphology is not understood yet.

Furthermore, some problems regarding the line emission of H$_2$ and neutral atoms in 
PNe are not solved yet. The excitation mechanism of the observed H$_2$ emission lines, for example, is still uncertain. Some authors conclude that excitation is due to radiative cascades following the UV pumping. Shock waves produced by the expanding nebulae are also suggested as a possible excitation mechanism. See, for example, Black, Porter \& Dalgarno (1981), Schild (1995), Kastner et al. (1996), Natta \& Hollenbach (1998), Hora et al. (1999), Guerrero et al. (2000), and Arias et al. (2001). Another problem is that the emission lines produced in low-ionization regions by neutral atoms like C, N, and O are not well fitted by the models (Aller 1984; Clegg et al. 1987; van Hoof et al. 2000; Stasinska \& Szczerba 2001). The analysis of these lines is made using photoionization models, where lines of neutral atoms are produced in the outer regions of a photoionized gas, and the presence of molecules is not taken into account. The hydrogen molecule, being the most abundant, can affect the physical conditions of the gas, since its presence changes the concentration of the different species in the gas, as well as the thermal equilibrium, due to the mechanisms involving H$_2$. If molecules are indeed present in the region where lines of neutral C, N, and O are produced, and if their effect on the gas is not taken into account, the results can be misleading. A detailed study, with a self-consistent modeling of the presence of H$_2$ molecules, is then important to a better knowledge of the chemical composition and physical conditions of PNe.

The aim of this paper is to make a first estimate of the amount of the H$_2$ molecule in the ionized regions of planetary nebulae, in order to verify if the presence of this molecule can be significant in these regions, as suggested by the observations. If the calculations show that the amount of H$_2$ is important, a next step would be a careful calculation of the level population distribution, in order to properly account for the different mechanisms that depend on the level population. In the present paper, the equations for the chemical equilibrium of molecular hydrogen are coupled to the equations of a photoionization code. The concentration of the hydrogen molecule is obtained along the nebula, as well as the total amount of molecules inside it. The theoretical models used in our analysis are described in \S~\ref{models}. The results for a standard PN and the effect of the input parameters on the obtained H$_2$ mass appear in \S~\ref{results}, while a discussion of the main results is presented in \S~\ref{conclusions}.

\section{MODELS} \label{models}

Models for typical PNe are obtained with the photoionization code AANGABA (Gruenwald \& Viegas 1992). The physical conditions of the gas are determined by solving the coupled equations of ionization and thermal balance for a spherically symmetric cloud. Several processes of ionization and recombination, as well as of gas heating and cooling, are taken into account. The energy input to the gas is essentially due to ionization of atoms and atomic ions by primary (stellar) and secondary (diffuse) radiation. The processes contributing to the gas cooling are colisional excitation, radiative and dielectronic recombination, thermal ionization and free-free emission. The transfer of the primary and diffuse radiation fields is treated in the outward-only approximation. Twelve elements (H, He, C, N, O, Mg, Ne, Si, S, Ar, Cl, and Fe) are included. The code was modified in order to include the chemical equilibrium of the species H, H$^+$, H$^-$, H$_2$, H$_2^+$, and H$_3^+$. The corresponding equations were then coupled with the equations for the ionization balance for the atomic gas. The H$_3$ molecule is not included due to its short lifetime ($\sim 10^{-7}$s; Gaillard et al. 1983; Bordas, Cosby, \& Helm 1990). The thermal balance was solved for the atoms and atomic ions; the effect due to H$^-$, H$_2$, H$_2^+$, and H$_3^+$ on the thermal balance is not taken into account.

Several processes that can form or destroy the included species are taken into account, as for example chemical reactions, photoprocesses, as well as grain surface reaction. The forty reactions included in the hydrogen equilibrium are listed in Table 1. Rate coefficients and cross sections are taken from the literature (see references in Table 1). Special attention was given to the search of the reactions and their corresponding rate coefficients, in order to have a set as complete and accurate as possible for the physical conditions prevailing in PNe. Some rate coefficients depend on the distribution of the reactants on their different energy levels. As said above, we do not calculate in detail the level distribution of the molecules in the present paper; a LTE level distribution for the hydrogen molecules, taking into account the three lowest vibrational levels of the electronic ground state is, however, adopted when cross-sections for the photoprocess are provided by the literature for individual levels.

Grains are assumed to be uniformly mixed with the gas, and are heated by the stellar and nebular continuum radiation. The attenuation of the stellar and diffuse radiation by dust is taken into account; the effect of scattering by dust on the radiation field is, however, not included.

\subsection{The H$_2$ Formation on Grain Surface}

The formation of H$_2$ on grain surface requires an analysis of the grain survival. We assume that the grain can survive if the grain temperature ($T_g$) is smaller than the sublimation temperature, which is a characteristic of the grain material.

The grain temperature is calculated assuming radiative equilibrium. The grains are heated by the stellar and the nebular continuum radiation, and emit as blackbodies.

Following Tielens \& Hollenbach (1985) and Hollenbach \& Salpeter (1971), the rate of 
H$_2$ formation on grain surface, $\Gamma _g$, is given by the product of the rate of H-grain collisions by probabilistic factors that depend on the grain surface properties and on the gas temperature, i.e.,
\begin{equation}
\Gamma_g = \frac{Sf}{2} \pi a_g^2 v_Hn_gn(H)
\end{equation}
where $a_g$ is the grain radius, $v_H$ is the thermal velocity of the H atoms, $n_g$ is the grain density, and $n(H)$ is the neutral atomic H density. The probabilistic factors $S$ and $f$ are given by Hollenbach \& McKee (1979). These factors decrease with the grain temperature ($T_g$) and the gas temperature ($T$), particularly when $T_g$ is higher than 100K. They also depend on the properties of the grain material.

A new model for the H$_2$ formation in grains is suggested by Cazaux \& Tielens (2002). In this model, they include the "physiobsorption" of the H atom (which dominates in temperatures lower than 50-100K), as well as the "chemisorption", which would be more important for higher temperatures. However, as stated by Glover (2003), the subject is still not clear.

From equation (1), the rate of H$_2$ formation on grain surface is proportional to the total available surface area of all grains, i.e., $\Gamma _g \propto a_g^2 n_g$. However, following the most common notation in the literature, we chose to describe the quantity of grains by means of the dust-to-gas ratio,
\begin{equation}
\frac{M_d}{M_g}=\frac {4 \pi a_g^3 \rho_g n_g}{3m_H n_H (1+4n_{He}/n_H)},
\end{equation}where $m_H$ is the atomic mass of H, $n_H$ and $n_{He}$ are the total numerical density of hydrogen and helium nuclei, respectively, and $\rho_g$ is the grain specific density. Assuming a Maxwellian distribution of velocity for the H atoms, equation (1) can then be rewritten as
\begin{equation}
\Gamma _g \propto \frac{SfT^{-1/2}M_d/M_g n_Hn(H)}{\rho_ga_g},\label{eqgrain}
\end{equation}
We define for convenience $(M_d/M_g)a_g^{-1}$ hereafter as $\beta$.

\subsection{Model Input Parameters}

A grid of theoretical models is obtained with input parameters typical of PNe, which includes the incident spectrum, the gas density and the elemental abundances of the nebula, as well as the dust material, size, and density.

In our models, a blackbody radiation is assumed for the PN central star, characterized by its temperature ($T_{*}$) and luminosity ($L_{*}$). According to Pottasch (1984), $T_{*}$ ranges from 3$\times$10$^4$ to 5$\times$10$^5$ K, while $L_{*}$ ranges from 10$^2 L_{\sun}$ to 10$^4 L_{\sun}$. 

The nebular gas is characterized by its total hydrogen nuclei density ($n_H$), which is assumed constant inside the nebula, and by its elemental abundances. Gas densities typically range from 10$^2$ to 10$^5$ cm$^{-3}$ (Pottasch, 1984). We assume constant elemental abundances for the atomic gas along the nebulae, with the same values for all models. These are mean values for planetary nebulae taken from Kingsbourgh \& Barlow (1994), for He, C, N, O, Ne, S, and Ar, obtained from atomic line-emission intensities. For the elements whose abundance is not provided by these authors (Mg, Si, Cl, and Fe), we adopted the values given by Stasinska \& Tylenda (1986), in order to make a rough correction for grain depletion. Since the effect of these elements is not very effective for the gas cooling, their exact proportion in the form of grains will not be important. 

Typical grain radius ranges from 10$^{-3}\micron$ to 10 $\micron$ and dust-to-gas mass ratio from 10$^{-5}$ to 10$^{-1}$ in PNe (Stasinska \& Szczerba 1999). The composition of the ISM grains (Aller 1984; Evans 1994), i.e., silicate and carbon-based compounds, is assumed here for the PNe grains. In this work, we analyze the presence of amorphous carbon, graphite, silicate, smoothed ultraviolet silicate, silicon carbide (SiC), as well as neutral and ionized polycyclic aromatic hydrocarbons (PAHs). Optical properties for these astrophysical grains are taken from Rowan-Robinson (1986), Draine \& Lee (1984), Laor \& Draine (1993), Weingartner \& Draine (2001), and Li \& Draine (2001). These properties depend on the grain size and composition, as well as on the photon energy. Sublimation temperatures and specific densities are taken from Laor \& Draine (1993) and Li \& Draine (2000).

We define a standard PN with a given set of input parameters, in order to explore the influence due to each one on the produced amount of H$_2$. For this, we vary each parameter along its typical range, while keeping the other parameters fixed. The input parameters for the standard PN are the following: $T_{*} =$ 150000 K, $L_{*} = 3000 L_{\sun}$, $n_H = 10^3$ cm$^{-3}$, $a_g = 0.1\mu$m, $M_d/M_g = 10^{-3}$, and amorphous carbon grains. 

\section{RESULTS} \label{results}

\subsection{Standard Model}

The calculated concentration of species containing hydrogen nuclei versus the distance from the central star is shown on the upper panel of Figure 1, for the standard PN model. The concentrations are given in units of $n_H$. The temperature radial profile for the standard nebula is plotted on the lower panel. The distance is given in units of the outer radius of the ionized region, $R_{max}$, which is defined as the distance to the central star where the ionization degree, $n(H^+)/n_H$, reaches 10$^{-4}$. For the standard PN, $R_{max}=$ 0.21 pc; for a nebula with same input parameters, but $T_*=$ 50000 K, $R_{max}=$ 0.17 pc, and, if the nebula has a density $n_H = 10^4$ cm$^{-3}$ and other parameters as for the standard nebula, $R_{max} = 0.04$ pc.

As it can be seen in Figure 1, the hydrogen molecule is mainly ionized in the inner and intermediate regions. The abundance of the neutral molecule, as well as of the ionized molecule, has its maximum in the inner region of the recombination zone, decreasing slightly in the less ionized regions. This behavior can be explained by the curves of Figure 2, which shows the rates for the most important reactions of formation and destruction of the hydrogen molecule in the standard PN, as a function of the distance from the central star. 

In brief, the most important reactions for the formation of the hydrogen molecule nuclei (H$_2$ or H$_2^+$) are:
\begin{mathletters} 
\begin{eqnarray} 
H + H^- \rightarrow H_2 + e^-\\ 
2H + grain \rightarrow H_2 + grain\\ 
H + H^+ \rightarrow H_2^+ + h\nu\\ 
H^+ + H^- \rightarrow H_2^+ + e^-
\end{eqnarray} 
\end{mathletters}with $H^-$ produced by the reaction
\begin{mathletters} 
\begin{eqnarray} 
H + e^- \rightarrow H^- + h\nu
\end{eqnarray}
\end{mathletters}Charge exchange reactions are effective in ionizing or recombining the hydrogen molecule:
\begin{mathletters} 
\begin{eqnarray} 
H_2^+ + H \rightarrow H_2 + H^+\\ 
H_2 + H^+ \rightarrow H_2^++H\\
\end{eqnarray} 
\end{mathletters}

Comparing the distribution of the various species (Fig. 1) with the formation and destruction rates along the standard PN (Fig. 2), the maximum concentration of molecular nuclei can be associated to a semi-ionized region, where H, H$^+$ and e$^-$ coexist with significant abundance. In fact, as said above, the maximum abundance of molecular hydrogen occurs in the inner recombination zone, and the distribution follows a radial distribution similar to that of H$^-$.

Grains are the most important route to the formation of H$_2$ in the outer regions since other reactions include ionized species. Therefore, when grains are not present the concentration of the molecule decreases in the outer regions. 

Regarding the destruction mechanisms, photodissociation is an important process for the destruction of H$_2$, as well as of H$_2^+$. In particular, the photodissociation in two steps is important for the destruction of H$_2$ in the whole nebula, since the maximum threshold for this process is lower than that of hydrogen. The direct photodissociation, as well as photoionization of H$_2$ are of minor importance, mainly in the intermediate and outer zones, since the energy threshold is higher than 13.6 eV. 

The abundance of H$_3^+$ is much lower compared with that of other species, and has its maximum in the same region as H$_2$ and H$_2^+$. The formation of this molecular ion occurs mainly through radiative association of H$_2$ with H$^+$, but the ion-molecule reaction of H$_2$ and H$_2^+$ is also important in the recombination zone. Dissociative recombination and collisions with H are the main destruction routes of H$_3^+$.

\subsection {The Effect of Dust}

The grain temperature depends on the incident radiation, as well as on the grain optical properties, which, in turn, depend on the grain material and size. The grain temperature decreases steeply in the inner 20\% region of the nebula, and decreases slowly for increasing radius. For example, assuming a$_g$ $\leq$ 0.01$\mu$m, and all other input parameters the same as those for the standard model, the temperature for graphite grains decreases from $\sim$ 500 K, in the inner regions, to $\sim$ 50 K, in the farthest zones, while for a$_g$ $\leq$ 0.1$\mu$m, the grain temperature ranges from 180 K to 26 K. For other materials the temperatures are slightly lower. Grains can survive inside the entire ionized region, except those made of PAH, which sublimate for distances shorter than $\sim$ 0.01$R_{max}$. 

Direct stellar radiation is much more important for the grain heating than absorption of Ly$\alpha$ radiation in the ionized region. The grain temperature is increased by $\sim$ 10\% when Ly$\alpha$ radiation is included in the grain heating. Such increase leads to a maximum decrease of only 0.3\% in the H$_2$ concentration.

The effect of the dust-to-gas ratio on the relative abundance of H$_2$ molecules is shown in Figure 3. The results presented in this figure correspond to models with the same input parameters as in the standard model, except for the dust-to-gas ratio. In the figure, the behavior of $n(H_2)$ with $R/R_{max}$ is shown for different values of $M_d/M_g$, with the lowest curve corresponding to a model without dust. Since the reaction on grain surface is an important reaction of H$_2$ formation all along the nebula, increasing $M_d/M_g$ results in higher values of $n(H_2)/n_H$. It can be also noticed that for $R/R_{max}$ between 0.8 and 0.9 the results do not depend on $M_d/M_g$. This can be explained by the fact that in this region the dominant processes of H$_2$ formation are the charge exchange and the associative detachment reactions. As said above, when few or no grains are present, there is a steep decrease of H$_2$ molecules in the outer regions of the nebula, due to the decrease of H$^-$, which is mainly produced in regions where the electron density is significant. 

As discussed in \S 2.1, the key parameter for the formation of H$_2$ in the surface of grains is the total available surface area of grains or, in terms of our input parameters, the dust-to-gas mass ratio divided by the grain radius ($\beta$). For a given dust-to-gas ratio, smaller grains imply a higher formation rate (see eq.[~\ref{eqgrain}]) of H$_2$ molecules. On the other hand, smaller grains have larger temperatures, leading to lower formation rates. However, the effect due to the grain temperature is small; fixing the dust-to-gas ratio, a decrease of the size of the grain leads to an increase of the abundance of H$_2$. 

As said in \S 1, the formation of H$_2$ molecules on the surface of grains has been assumed in the interstellar medium in order to explain the required amount of H$_2$ molecules. For the ionized region of PNe, however, the associative detachment reaction, as well as the charge transfer of H$_2^+$ with H are also important processes for the formation of H$_2$ molecules. The dust contributes more significantly for the production of H$_2$ molecules for values of $\beta >$ 100 cm$^{-1}$.

\subsection {Other models}

The distribution of the various species, as well as that of the reactions rates, along the nebula for models with input parameters other than those of the standard model is similar to that shown in Figures 1 and 2. In all models, the maximum H$_2$ concentration occurs where the gas temperature is around 7500 K and the ionization degree $n(H^+)/n_H$ is approximately 0.1.

In order to compare the amount of H$_2$ produced in different PNe conditions, we define the parameter $R_M$, which corresponds to the ratio between the total hydrogen molecule nuclei mass (H$_2$ + H$_2^+$) and the total hydrogen nuclei mass, i.e., 
\begin{equation} 
R_M = \frac{\int_{R_0}^{R_{max}}4\pi 
[n(H_2)+n(H_2^+)]m_{H_2}r^2dr}{\int_{R_0}^{R_{max}}4\pi n_Hm_Hr^2dr}, 
\end{equation} where $r$ is the distance from the central star, while $R_0$ and 
$R_{max}$ are the inner and the outer radius of the ionized region as defined above, respectively. The inner radius is chosen as $R_0$ = 10$^{15}$ cm, but its exact value does not affect the results presented here, provided that it is lower than 0.3$R_{max}$. 

For the standard model, we obtain a value for $R_M$ of the order of 2$\times$10$^{-6}$. Increasing the dust-to-gas ratio (or decreasing the grain size) by a factor of 10, and keeping the other input parameters fixed, the fraction of molecular mass increases by a factor around 2. As discussed above, the main parameter regarding the presence of grains is the available total surface of grains. When dust is not present in the gas, $R_M$ is only 10\% lower than in the standard model. In fact, $R_M$ increases with $M_d/M_g$ and/or decreasing $a_g$, but, as said above, for a value of $\beta$ lower than that for the standard model, the presence of grains does not affect the obtained results for $R_M$. 

Different grain materials leads to a maximum difference in $R_M$ by less than 25\%.

Plots of $R_M$ versus $T_*$ are shown in Figure 4, for different values of $L_*$, $n_H$, and $M_d/M_g$. In this figure, it can be seen that $R_M$ increases with $T_*$. This is an interesting result, since hotter stars produce more high-energy photons. However, since the hydrogen photoionization cross section decreases with the photon energy, high-energy photons can travel farther into the nebula before being absorbed by the gas, producing a more extended recombination zone. The increase in $R_M$ with $T_*$ can explain why H$_2$ is more frequently detected in bipolar PNe, since these objects have hotter central stars (Corradi \& Schwarz 1995). In fact, the relative thickness of the recombination zone is the crucial factor for $R_M$, and it implies that $R_M$ also increases with the decrease of $n_H$ and $L_*$, as shown in Figure 4. An increase of the gas density by a factor 10 decreases $R_M$ by a factor of the order of 3, while a change of a factor 100 in the luminosity leads to a change of $R_M$ by a factor around 24. However, $R_M$ increases more steeply with $T_*$ than with any other input parameter, as can be seen from Figure 4. An increase of $T_*$ by a factor 10 leads to an increase of $R_M$ by a factor around 6000.

\section{CONCLUDING REMARKS} \label{conclusions}

Using a photoionization code which includes atoms and ions of twelve elements, as well as H$^-$, H$_2$, H$_2^+$, and H$_3^+$, we analyze the presence of the H$_2$ molecule over the entire ionized region of PNe, up to the region where the ionization degree decreases to 10$^{-4}$. The transfer of the radiation emitted by the star, as well as of that produced by the nebula, is treated in detail in order to obtain a self-consistent distribution of the hydrogen molecule in the nebula. The presence of grains is taken into account, since they are present in the ionized region and can act as catalyst in the formation of the H$_2$ molecule. The effect of the various reactions that contribute to the formation and destruction of H$_2$ is studied. 

Our calculations show that H$_2$ molecules can survive inside the ionized region of PNe, particularly in the inner zone of the recombination region, where the ionization degree is of the order of 0.1. In this region, the formation rate of H$_2$ molecules is enhanced due to the high efficiency of the associative detachment of H and H$^-$ and the charge exchange between H and H$_2^+$, while the destruction is not very effective due to the shielding provided by H atoms. When significant superficial area of grains is available (i.e., $\beta >100$ cm$^{-1}$), the grain reaction is the main process of H$_2$ production in regions of low ionization.

Notice that the grain surface reaction is not the only important production mechanism of H$_2$ molecules in PNe, as it is in the interstellar medium. The associative detachment of H and H$^-$, as well as the charge exchange between H and H$_2^+$ are also important processes for the formation of H$_2$ molecules. In the H recombination zone, where the density of H$_2$ reaches its higher values, the main processes, even when large amounts of grains are present, are via associative detachment and charge exchange. The H$^-$ and the H$_2^+$ ions have similar concentrations, except in the outer region of the PNe, and are abundant in regions where both electrons and H ions are present.

The effect of the stellar temperature and luminosity, as well as that of the gas density, dust-to-gas mass fraction, and grain material, on the obtained abundance of the hydrogen molecule, is also analyzed. Regarding the presence of grains, the available superficial area is the main parameter that controls the molecule formation efficiency, while the grain temperature and composition are of minor importance.

We conclude that $R_M$, the H$_2$ to total hydrogen nuclei mass ratio, increases with the relative thickness of the recombination zone. This explains why $R_M$ increases with $T_*$ and decreases with $n_H$ and $L_*$, the main effect being due to the stellar temperature. The behavior of $R_M$ versus $T_*$ can explain why the emission of H$_2$ is more often detected in PNe with bipolar morphology, since these objects have in general hotter central stars.

The presence of molecular material is generally used as evidence that the region is optically thick. However, if the hydrogen molecule can survive within the hot gas, the detection of H$_2$ emission is not an indication that a nebula is optically thick. This point has important consequences for the determination of PNe distances, since some distance indicators are obtained assuming that the gas is completely optically thick (i.e., all ionizing photons are absorbed). The same applies to the Zanstra method for the determination of the stellar temperature. 

As shown in this paper, H$_2$ molecules can survive not just in very low, but also in mild ionization regions inside PNe. Thus, a detailed study of the effect the presence of the hydrogen molecule on the gas thermal equilibrium is essential for an accurate picture of the physical processes inside PNe. The presence of H$_2$ molecules in mild ionized gas can solve, for example, some problems faced by photoionization models in explaining emission lines from the low-ionization regions. Hydrogen molecules can affect the physical conditions of the gas, providing the additional heating required to explain the discrepancies. A more realistic estimate for the gas abundance could also be obtained.

In this work we do not include the self-shielding of H$_2$ in the calculations, since it would be necessary to obtain the distribution of the molecule in its various energy levels in order to estimate the shielding. The effect of the self-shielding on the photoprocessess can be very important in the outer regions of PNe where the H$_2$ column density is high. Approximated methods for the treatment of the self-shielding effect on the photodissociation of H$_2$ are available in the literature as, for example, in Van der Werf (1989) and Draine \& Bertoldi (1996). Taking this approximated methods into account, the computed H$_2$ concentration can increase up to two orders of magnitude in the outer parts of the ionized region. Therefore the calculated H$_2$ concentrations in the present work can be considered as lower limits.

Shocks are frequently invoked in the literature for explaining the H$_2$ line emission since pure radiative excitation does not account for the observations. The presence of H$_2$ molecules in regions of higher excitation level, excited by UV radiation not fully absorbed by inner regions, could account for the line emission, without invoking shocks.

Molecular and atomic hydrogen masses for the Ring Nebula (NGC 6720) were derived from observations by Greenhouse, Hayward, \& Thronson (1988), resulting in a molecular to atomic gas ratio equal to $\sim 7\times10^{-5}$. Assuming the equilibrium between formation on grain surface and photodissociation, these authors conclude that chemical equilibrium does not account for the detected quantity of H$_2$. However, from our more detailed model and assuming the parameters for the central star and the gas given by Greenhouse et al. (1988), we obtain $R_M = 2.4\times10^{-5}$, which is reasonably similar to the value obtained by these authors.

\acknowledgments

We are thankful to S. M. Viegas for many useful comments, to E. Roueff for kindly providing the transitions probabilities and dissociation fractions for the hydrogen molecule, and to T. Abel for the helpful discussion about reaction rates. We are pleased to thank the anonymous referee for his valuable comments and suggestions that helped to improve the paper. The authors thank FAPESP, CNPq, and Pronex/CNPq for the financial support of this work.

\clearpage

\begin{figure}
\epsscale{0.75}
\plotone{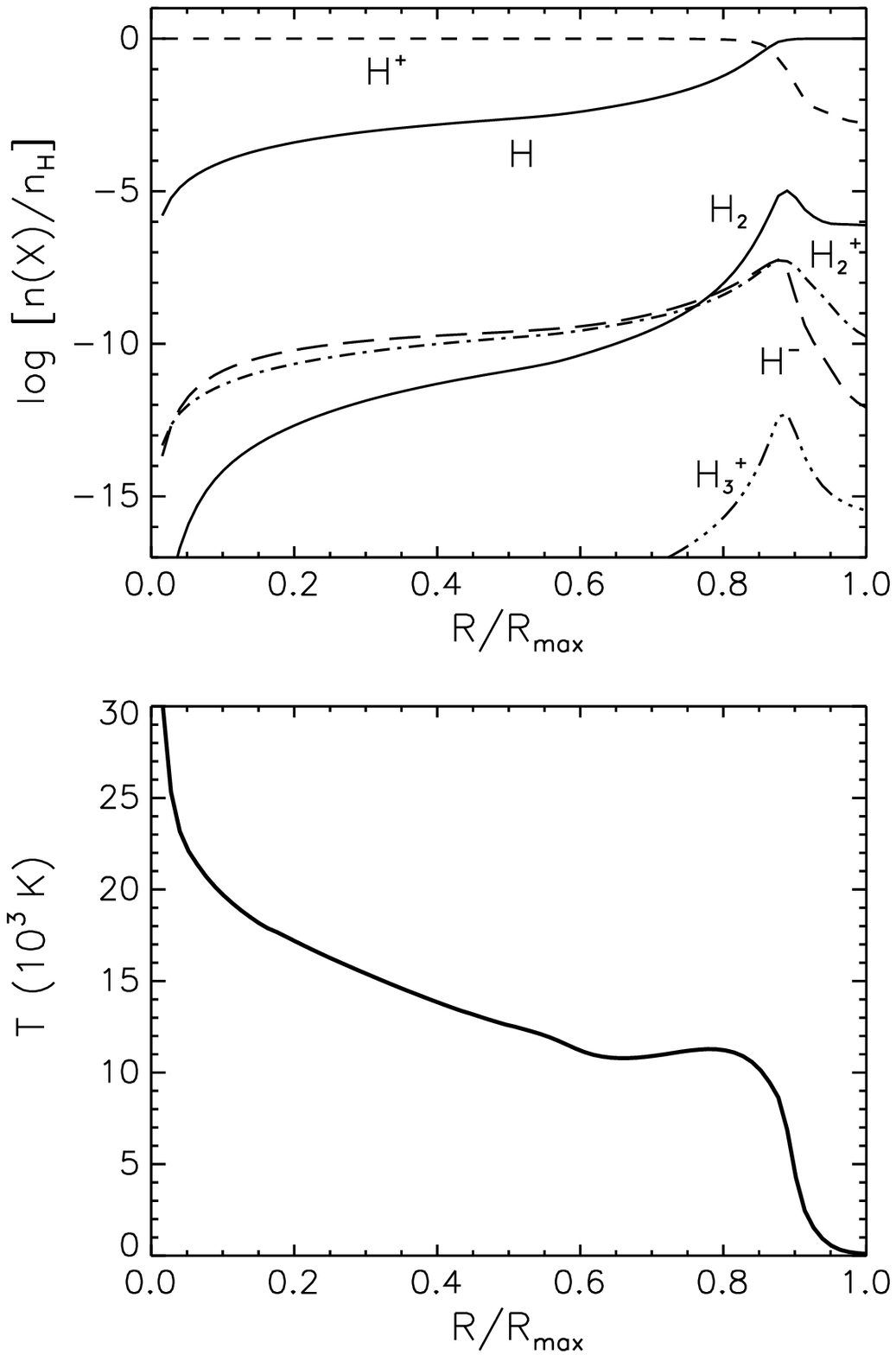}
\caption{Concentrations of the species containing H nuclei (upper panel) and gas temperature (lower panel), for the standard model. \label{fig1}}
\end{figure}


\begin{figure}
\epsscale{1}
\plotone{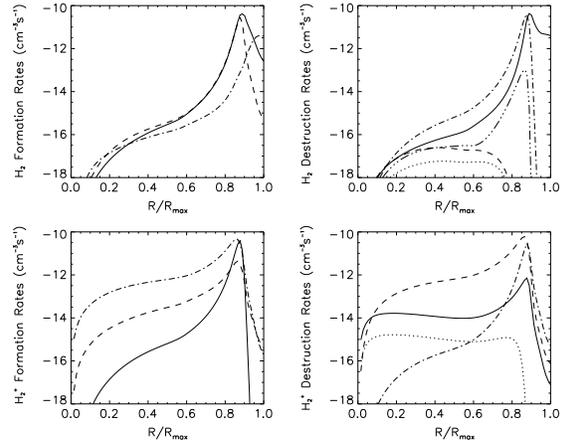}
\caption{Rates of the most important formation and destruction reactions of molecular hydrogen, for the standard model. H$_2$ formation (upper left): H + H$^-$ (solid curve), H$_2^+$ + H (dashed), and H + H + grain (dot-dashed). H$_2$ destruction (upper right): H$_2$ + h$\nu$ (photodissociation in two steps; solid), H$_2$ + h$\nu$ (direct photodissociation; dashed), H$_2$ + H$^+$ (charge exchange, dot-dashed), H$_2$ + e$^-$ (colisional dissociation, dot-dot-dot-dashed), and H$_2$ + h$\nu$ (photoionization; dotted). H$_2^+$ formation (lower left): H$_2$ + H$^+$ (solid), H$^-$ + H$^+$ (dashed), and H + H$^+$ (dot-dashed). H$_2^+$ destruction (lower right): H$_2^+$ + h$\nu$ (producing H$^+$ + H; solid), H$_2^+$ + e$^-$ (dashed), H$_2^+$ + H (dot-dashed), and H$_2^+$ + h$\nu$ (producing 2H$^+$ + e$^-$; dotted). \label{fig2}}\end{figure}


\begin{figure}
\plotone{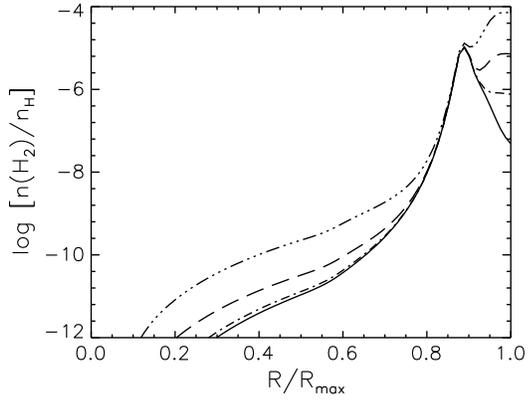}
\caption{Effect of the dust-to-gas mass ratio on the calculated abundance of H$_2$. 
Curves are shown for $M_d/M_g$ equal to 0, 10$^{-3}$, 10$^{-2}$, and 10$^{-1}$; the lower curve corresponds to a model without grains, while upper curves correspond to models with higher values of the dust-to-gas ratio. Other parameters are the same as in the standard PN model. \label{fig3}}
\end{figure}


\begin{figure}
\epsscale{0.75}
\plotone{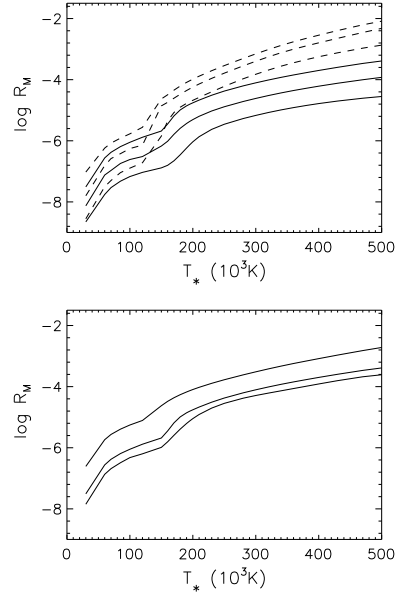}
\caption{Molecular hydrogen mass fraction versus central stellar temperature. 
Upper panel: The results are shown for $n_H = 10^3$, 10$^4$, and 10$^5$ cm$^{-3}$ (lower curves correspond to higher $n_H$) for two values of $\beta$: 10$^{2}$ cm$^{-1}$ (solid curves) and 10$^{4}$ cm$^{-1}$ (dashed). Lower panel: Results for $L_*$ = 500, 3000, and 10$^4 L_{\sun}$, with $R_M$ decreasing with increasing $L_*$. Other parameters correspond to those of the standard model. \label{fig4}}
\end{figure}

\clearpage 

\begin{deluxetable}{lclc}
\tablewidth{0pt}
\tablecaption{Chemical Reactions and Physical Processes}
\tablehead{\colhead{Reaction}&\colhead{Ref.}&\colhead{Reaction}&\colhead{Ref.}}
\startdata

Photoionization&&Photodissociation&\\
$H+h\nu\rightarrow H^++e^-$&1&$H_2+h\nu\rightarrow 2H$&12\\
$H_2+h\nu\rightarrow H_2^++e^-$&2&$H_2+h\nu \rightarrow H_2^\ast\rightarrow 2H$&13\\
&&$H_2^++h\nu \rightarrow H+H^+$&14\\
Photodetachment&&$H_2^++h\nu \rightarrow 2H^++e^-$&15\\
$H^-+h\nu \rightarrow H+e^-$&3&&\\
&&Radiative Attachment&\\
Radiative Recombination&&$H+e^-\rightarrow H^-+h\nu$&16\\ 
$H^++e^-\rightarrow H+h\nu$&4&&\\
&&Charge Exchange&\\
Dissociative Recombination&&$H_2^++H\rightarrow H^++H_2$&17\\
$H_2^++e^-\rightarrow 2H$&5&$H_2+H^+\rightarrow H+H_2^+$&7\\
$H_3^++e^-\rightarrow H_2+H$&6&$He^++H\rightarrow He+H^+$&18\\
$H_3^++e^-\rightarrow 3H$&6&$He^{++} + H \rightarrow He^+ + H^+$&19\\
&&&\\
Associative Detachment&&Radiative Association&\\
$H + H^- \rightarrow H_2 + e^-$&7&$H + H^+ \rightarrow H_2^+ + h\nu$&7\\
$H^+ + H^- \rightarrow H_2^+ + e^-$&7&$H_2 + H^+ \rightarrow H_3^+ + h\nu$&7\\
&&&\\
Ion-Molecule Reaction&&Grain Surface Reaction&\\
$H_2^+ + H_2 \rightarrow H_3^+ + H$&8&$2H + grain \rightarrow H_2 + grain$&20\\
&&&\\
Neutralization&&Other Processes&\\
$H^+ + H^- \rightarrow 2H$&9&$H_2 + e^- \rightarrow 2H + e^-$&21\\
$H_2^+ + H^- \rightarrow H_2 + H$&9, 10&$H_2 + e^- \rightarrow H + H^-$&22\\
$H_2^+ + H^- \rightarrow 3H$&9, 10&$H_2 + H \rightarrow 3H$&23\\
$H_3^+ + H^- \rightarrow 2H_2$&9, 10&$2H_2 \rightarrow H_2 + 2H$&23\\
$H_3^+ + H^- \rightarrow H_2 + 2H$&9, 10&$H^- + e^- \rightarrow H + 2e^-$&16\\
&&$H^- + H \rightarrow 2H + e^-$&22\\
Colisional Ionization&&$H_2^+ + H_2 \rightarrow H + H^+ + H_2$&11\\
$H + e^- \rightarrow H^+ + 2e^-$&9&$H_3^+ + H \rightarrow H_2 + H_2^+$&7\\
$2H \rightarrow H + H^+ + e^-$&11&$H_3^+ + H_2 \rightarrow H + H_2 + H_2^+$&11\\
$2H \rightarrow H + H^{\ast} \rightarrow H + H^+ + e^-$&11&$H_3^+ + H_2 \rightarrow H^+ + 2H_2$&11\\
\enddata
\tablerefs{(1) Osterbrook (1989); (2) Yan, Sadeghpour, \& Dalgarno (1998, 2001); (3) de Jong (1972); (4) P\'{e}quignot, Petitjean, \& Boisson (1991); (5) Schneider et al. (1994, 1997); (6) McCall et al. (2003), with branching ratios given by Datz et al. (1995); (7) Galli \& Palla (1998); (8) Theard \& Huntress (1974); (9) Lepp, Stancil, \& Dalgarno (2002); (10) following Dalgarno \& McCray (1973) we assume this rate equal to that of the reaction H$^+ +$ H$^- \rightarrow$ 2H; (11) Hollenbach \& McKee (1989); (12) Allison \& Dalgarno (1969), with LTE distribution for the H$_2$ energy levels; (13) Abgrall et al. (1994) and Abgrall, Roueff, \& Drira (2000), with LTE distribution for the H$_2$ energy levels; (14) Stancil (1994), with LTE fitting to T $=$ 8400 K; (15) Bates \& Opick (1968); (16) Abel et al. (1997); (17) Karpas, Anicich, \& Huntress (1979); (18) Jura \& Dalgarno (1971); (19) Arthurs \& Hyslop (1957); (20) see text; (21) Donahue \& Shull (1991); (22) Izotov \& Kolesnik (1984); (23) Shapiro \& Kang (1987).}
\end{deluxetable}
\end{document}